\documentstyle[11pt]{article}   
\input epsf.sty
\title{\Large \bf
Input-output behaviour of a model neuron \\
with alternating drift
}
\author{
 {\sc \small{A.BUONOCORE$^{(1)}$}}
 {\sc  \small{A. DI CRESCENZO$^{(2)}$\footnote{corresponding author}}, }
 {\sc \small{E. DI NARDO$^{(3)}$}, }\\
 \\
 {\small \rm (1) \ Dipartimento di Matematica e Applicazioni, 
 Universit\`a di Napoli Federico II}\\
 {\small \rm Via Cintia, I-80126 Napoli, Italy}\\
 {\small\rm E-mail: \tt aniello.buonocore@unina.it}\\
 \\
 {\small \rm (2) \ Dipartimento di Matematica e Informatica,
 Universit\`a di Salerno} \\
 {\small \rm Via Ponte don Melillo, I-84084 Fisciano (SA), Italy} \\
 {\small \rm E-mail: \tt adicrescenzo@unisa.it}\\
 \\
 {\small \rm (3) \ Dipartimento di Matematica,
 Universit\`a degli Studi della Basilicata}\\
 {\small \rm C.da\ Macchia Romana, I-85100 Potenza, Italy}\\
 {\small \rm E-mail: \tt dinardo@unibas.it}\\
}
\date{}
\begin{document}
\setlength{\baselineskip}{13pt}
\maketitle
\begin{abstract}
The input-output behaviour of the Wiener neuronal model subject to alternating
input is studied under the assumption that the effect of such an input is
to make the drift itself of an alternating type. Firing densities and related
statistics are obtained via simulations of the sample-paths of the process
in the following three cases: the drift changes occur during random periods 
characterized by {\em (i)} exponential distribution, {\em (ii)} Erlang 
distribution with a preassigned shape parameter, and {\em (iii)} deterministic 
distribution. The obtained results are compared with those holding 
for the Wiener neuronal model subject to sinusoidal input.
\end{abstract}

\medskip\noindent
{\bf Keywords:} Wiener neuronal model; firing densities;
alternating drift.

\section{Introduction}\label{section:1}
During the last four decades numerous efforts have been devoted to the construction of mathematical
models aiming at the description of single neuron's firing processes. A customary feature of
widespread existing models is the assumption that the neurons' firing is described by the first
passage of the membrane potential through a threshold, the membrane potential being
viewed as a continuous stochastic process. Great attention has being put on the identification of
suitable Markov processes and firing thresholds, especially with a view to include in the model
certain relevant features, such as the effects of external input on the neuron's response
(see Ricciardi, 1995, and Ricciardi {\em et al.}, 1999, for a description of neuronal models
and methods to face the first-passage-time problem).
\par
Within such a background, in this paper we study the input-output behaviour of model neurons 
subject to alternating input, assuming that excitatory stimuli prevail on inhibititory 
stimuli, and viceversa, during randomly alternating time intervals. The membrane potential 
is assumed to be described by a Wiener process; the effects of the alternating stimuli is 
condensed in the drift. The first passage through constant firing thresholds is analyzed
by constructing estimated firing densities. This is obtained by making use of an ad hoc
procedure for the simulation of sample paths of the Wiener process with alternating drift.
The effect of alternating stimuli on the firing densities is
analyzed when the drift changes occur during random periods characterized by 
{\em (i)\/} exponential distribution, {\em (ii)\/} Erlang distribution with 
shape parameter 2, and {\em (iii)\/} deterministic distribution.
The computational results obtained for the considered model are then compared with
those holding for the Wiener neuronal model subject to sinusoidal input.
\par
Let us now point out some specific features of this paper with a view to relate them 
to other works in the field. First of all, as mentioned in L\'ansk\'y et al.\ (2001), 
the interpretation of the neuronal firing times in terms of alternating periods 
appears to be natural for some types of neurons. It should be also mentioned that 
a similar problem 
was analyzed in Bulsara et al.\ (1994), where the first passage time problem of 
a Wiener process with sinusoidal time-varying drift through a constant threshold 
is treated. Our proposed model, however, offers the advantages that the epochs of 
high and low stimulation can be of different lengths; moreover, due to its greater 
flexibility, it allows to describe time-varying inputs characterized by more 
general shapes than sinusoidal ones. 
\section{The Wiener model}\label{section:2}
As customary, we assume that the neuron's membrane potential is described by a
continuous one-dimensional stochastic process $\{X(t);\,t\geq 0\}$ representing the changes
in the membrane potential between two consecutive neuronal firings, the firing threshold
being a constant $\beta$. Assuming that $X(0)=x_0<\beta$, the FPT random variable
$$
 T=\inf\{t\geq 0\colon\,X(t)>\beta\}
$$
describes the neuronal interspike intervals. An essential problem is
to determine the p.d.f.\ of $T$, i.e.\ the firing density
$$
 g(\beta,t\,|\,x_0)={\partial\over\partial t}\,\Pr(T\leq t), \quad t> 0.
$$
\par
A first well-known neuronal model was proposed by Gerstein and Mandelbrot~(1964),
who assumed that the neuron's membrane potential undergoes a simple random 
walk under the effect of excitatory and inhibitory synaptic actions. 
This assumption, upon a suitable limit procedure, leads to the Wiener process 
$$
 X(t)=x_0+\xi\,t+\sigma\,B(t), \quad t> 0,
$$
with $x_0\in{\bf R}$, $\xi\in{\bf R}$ and $\sigma>0$, and
where $B(t)$ denotes  the standard Brownian motion. Even though
the assumptions of this model are  oversimplified as they do
not take into account certain properties such as the
spontaneous  exponential decay of the membrane potential in the
absence of input, the Wiener model  can be usefully treated as
a starting point for investigations on neuronal stochastic
models.  For instance, the Wiener model with random jumps has
been considered recently by  Giraudo et al.\ (2001) as a model
for the description of changes in the membrane  potential
depending on the distance between the trigger zone and the
synaptic ending.
\section{Wiener model with alternating drift}\label{section:3}
Our present aim is to analyse the input-output behaviour of the Wiener model
subject to alternating input. To such a purpose, we assume that the 
effects of alternating stimuli reflect into the drift of the Wiener process. 
This leads to a Wiener process with alternating drift, described by 
\begin{equation}
 X(t)=x_0+\int_0^t \xi(s)\,{\rm d}s+\sigma\,B(t), 
 \quad t\geq 0,
 \label{equation:1}
\end{equation}
where $x_0\in{\bf R}$, $\sigma>0$ and $\{\xi(t);\,t\geq 0\}$ is a dichotomous 
stochastic process on states $c$ and $-v$, with $c,v>0$ and $\xi(0)=c$. 
Eq.~(\ref{equation:1}) describes a Brownian motion that starts from $x_0$ 
at time $0$, with positive drift $c$ and infinitesimal variance $\sigma^2$. 
Denoting by $Y_1,Y_2,\ldots$ the independent random times separating consecutive 
changes of the drift value, we have $\xi(t)=c$ during the random periods of 
lengths $Y_1,Y_3,Y_5,\ldots$ and $\xi(t)=-v$ during the remaining random 
periods. Formally, 
$$
 \xi(t)=\frac{c-v}{2}+\frac{c+v}{2}\,(-1)^{N(t)}, 
 \quad t\geq 0,
$$
where 
$$
 N(t)=\sum_{k=1}^{\infty}{\bf 1}_{\{Y_1+\cdots+Y_k\leq t\}}, 
 \quad t\geq 0.
$$
At the random times $Y_1+\cdots+Y_k$ the value of $\xi(t)$ thus 
changes from $c$ to $-v$ if $k$ is odd, and from $-v$ to $c$ if $k$ 
is even. We also assume that the non-negative 
random variables $Y_1,Y_3,Y_5,\ldots$ and $Y_2,Y_4,Y_6,\ldots$ have 
distribution functions $F_U(t)$ and $F_D(t)$, respectively, so that 
the random periods during which the drift is positive (resp.\ negative) 
are i.i.d. We remark that the probability law of~(\ref{equation:1}) can 
be expressed as a mixture of Gaussian densities (see Di~Crescenzo, 2000).  
\section{Estimating the firing density via simulations}\label{section:4}
In this Section we shall determine the firing density of the Wiener neuronal
model with alternating drift in the presence of a constant firing threshold $\beta$.
Making use of a direct analysis of the sample-paths of process~(\ref{equation:1}), 
it is possible to obtain a series expression of the firing density. Unfortunately, 
the series involves integrals of progressively increasing dimensionality, so that 
such an analytic result is unsuitable for practical purposes. Nevertheless, by 
truncation to the first few terms the following lower bound for the firing density 
is obtained: For all $t>0$ and $\beta>x_0$ we have 
$$
g(\beta,t\,|\,x_0) \geq [1-F_U(t)]\,\widetilde g_c(\beta,t\,|\,x_0) + \int_0^t\int_{-\infty}^{\beta}[1-F_D(t-u)]\,\widetilde g_{-v}(\beta,t-u\,|\,x) \alpha(x,u\,|\,x_0)\,{\rm d}x\,{\rm d}F_U(u),$$
where $\widetilde g_{\eta}(\beta,t\,|\,x_0)$ is the first-passage-time density 
of a Wiener process with drift $\eta$ throught $\beta$,  
$$
  \widetilde g_{\eta}(\beta,t\,|\,x_0)
 ={\beta-x_0\over \sqrt{2\pi\sigma^2 t^3}} 
 \exp\left\{-{(\beta-x_0-\eta t)^2\over 2\sigma^2 t}\right\},
$$
while $\alpha(x,t\,|\,x_0)$ is the $\beta$-avoiding density of a Wiener process 
with drift $c$,  
$$ \alpha(x,t\,|\,x_0) = {1\over \sqrt{2\pi\sigma^2 t}} 
 \exp\left\{-{(x-x_0-c\,t)^2\over 2\sigma^2 t}\right\}  
 \left[1-\exp\left\{-{2\over\sigma^2 t}\,(\beta-x)\,(\beta-x_0)\right\}\right].$$
\par
In order to obtain histograms as estimates of the firing densities, we resort 
to simulations of the sample paths of process~(\ref{equation:1}). This is 
performed by making use of a procedure that is strictly based on 
the properties of the underlying Wiener process. Such procedure simulates 
the sample-paths of process~(\ref{equation:1}) at the random 
times where the drift alternates.
\\ \smallskip\\
A sketch of the procedure for the simulation of $T$ is given below:
\\ \smallskip\\
{\small \tt
\underline{Begin Procedure} \\
\textsc{0.}$\;$ $x:=x_0$, $t:=0$, $\xi:=c$; \\
\textsc{1.}$\;$ generate an inversion instant $\tau$; \\
\textsc{2.}$\;$ $p:= \Pr\{$the first passage occurs
before $\tau\}$; \\
\textsc{3.}$\;$ generate an uniform pseudo-random
number $u$ in $(0,1)$; \\
\textsc{4.}$\;$ if $u>p$ then goto \textsc{Step~7}; \\
\textsc{5.}$\;$ generate a pseudo-random number $\theta$
in $(0,\tau)$ from p.d.f.\ $f_1(\theta)$; \\
\textsc{6.}$\;$ $fpt:=\theta+t$; output($fpt$); stop; \\
\textsc{7.}$\;$ generate a pseudo-random number $z$
in $(-\infty,\beta)$ from p.d.f.\ $f_2(z)$; \\
\textsc{8.}$\;$ $x:=z$, $t:=t+\tau$; update $\xi$; \\
\textsc{9.}$\;$ goto \textsc{Step~1}; \\
\underline{End Procedure}
}
\\ \smallskip \\
For \textsc{Step 1} we assume that the random inversion times having distributions 
$F_U$ and $F_D$ can be numerically simulated.
\par
Let us informally discuss the underlying ideas of this procedure. After the first
drift inversion time $\tau$ has been generated, a Bernoulli trial with success probability
$p$ is simulated (\textsc{Step 2}), where $p$ is the FPT probability
$$ p:=P(T\leq \tau)=\Phi\left(-\displaystyle{\beta-x-\xi \tau\over \sigma\sqrt{\tau}}\right)  + e^{2\xi(\beta-x)/\sigma^2}\,
 \Phi\left(-\displaystyle{\beta-x+\xi \tau\over \sigma\sqrt{\tau}}\right),$$
$\Phi$ denoting the standard normal distribution function. A success of the Bernoulli 
trial means that the first passage has occurred before time $\tau$. In such a case the 
FPT is simulated via $f_1$, which is the FPT density of a Wiener process through $\beta$ 
conditional on $\{T\leq \tau\}$, and the simulation ends. If a failure occurs in the 
Bernoulli trial, then the value attained by the Wiener process at time $\tau$ is 
simulated via density $f_2$, which is the density of $X(\tau)$ conditional on 
$\{T>\tau\}$. Densities $f_1$ and $f_2$, whose analytic expressions are known, 
are simulated by making use of a standard Von Neumann acceptance-rejection method 
(see for instance Rubinstein, 1981). After $X(\tau)$ has been generated, the current 
drift $\eta$ is updated to the new value, i.e.\ $-v$ if it was $c$, and vice-versa. 
The simulation than proceeds, restarting afresh by generating a new inversion time 
of the drift, and so on until the first passage through $\beta$ occurs.
\par
We point out that, according to the assumptions of our model, the case of endogenous 
periodicity is developed in the given procedure. In other words, the input is reset 
after that the threshold is reached. However, we point out that the case of 
exhogenous periodicity could be considered upon an easy modification of the 
procedure (useful references on the effects of endogenous and exhogenous 
periodicity are the papers by L\'ansk\'y (1997) and Plesser and Geisel (2001)). 
\par
In the following sections we discuss the results obtained in three special cases 
for the drift inversion times: {\em (i)\/} exponentially distributed times, 
{\em (ii)\/} Erlang distributed times, and {\em (iii)\/} deterministic times. 
The shown estimates of the firing density $g(\beta,t\,|\,x_0)$ are obtained
via simulation of $10^6$ sample-paths of $X(t)$. 
\section{Exponentially distributed inversion times}\label{section:5}
Let us assume that the inversion times of the Wiener neuronal 
model~(\ref{equation:1}) have distribution functions
$F_U(t)=1-e^{-\lambda t}$ and $F_D(t)=1-e^{-\mu t}$, $t\geq 0$.
Thus, $\lambda^{-1}$ is the mean of the random periods during which
the excitatory stimuli prevail on inhibitory stimuli, while $\mu^{-1}$
is the mean of the random periods with opposite behaviour.
We have considered some cases of interest in the physiological contexts, 
with $\mu\geq\lambda$, with threshold $\beta=10$, and with $x_0=0$.
\par
The most evident feature of the estimated firing densities is their being 
unimodal (see Figure 1). In particular, if $\mu$ increases the mode $m$ 
increases while the median $q_2$ decreases, as well as the quartiles $q_1$ 
and $q_3$, and the inter-quartile range $IQR$. This is in agreement with 
the fact that if $\mu$ increases, the random periods with prevailing 
inhibitory stimuli becomes `smaller' in stochastic sense. A quite similar 
behaviour is observed if $\lambda$ decreases (see Figure 2). Moreover, as 
$\sigma$ increases the firing density becomes less peaked, and $q_1,q_2,q_3$ 
and $m$ decrease (see Figure 3). Finally, when $c$ and $v$ increase and $c=v$, 
then $q_1,q_2,q_3$ and $m$ decrease (see Figure 4), as it should be expected 
since for high values of $c$ the sample-paths of $X(t)$ are more likely 
to reach the firing threshold.

\begin{figure} 
\centering{
\epsfxsize=7.5cm
\epsfysize=7.5cm
\epsfbox{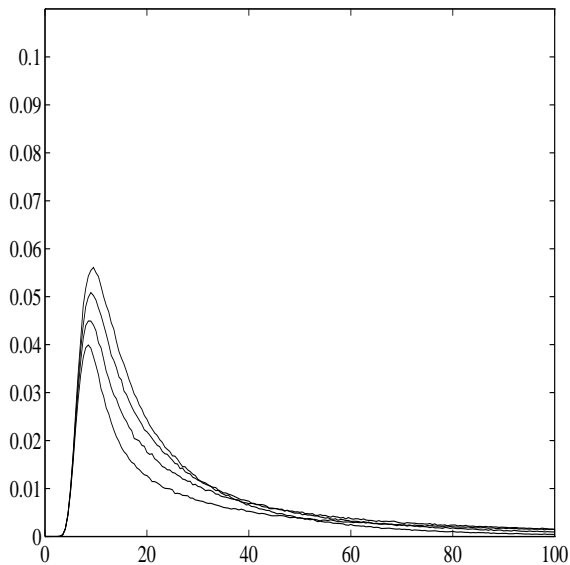}
\caption{\small Estimated firing density of the Wiener model with
exponentially alternating drift for $\sigma=1$, $\lambda=0.2$, $c=v=1$, with the choices $\mu=0.2, 0.3, 0.4, 0.5$ (bottom to top near the origin).}
}
\end{figure}

%
%
\begin{figure} 
\centering{
\epsfxsize=7.5cm
\epsfysize=7.5cm
\epsfbox{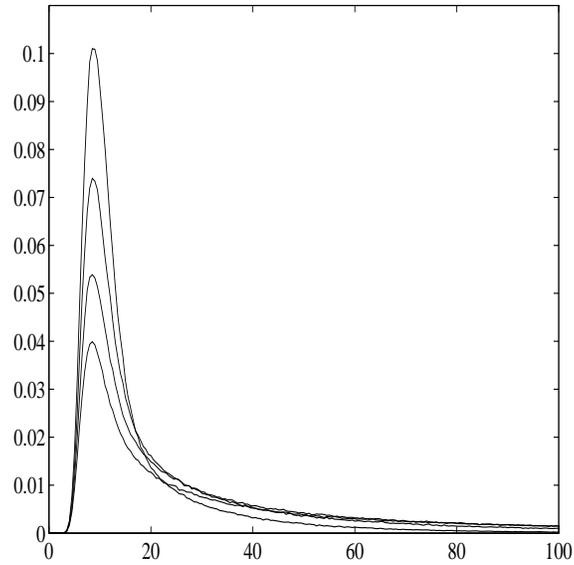}
\caption{\small As in Figure 1, with $\sigma=1$, $\mu=0.2$, $c=v=1$,
for $\lambda=0.2, 0.15, 0.1, 0.05$ (bottom to top near the origin).}
}
\end{figure}

%
\begin{figure} 
\centering{
\epsfxsize=7.5cm
\epsfysize=7.5cm
\epsfbox{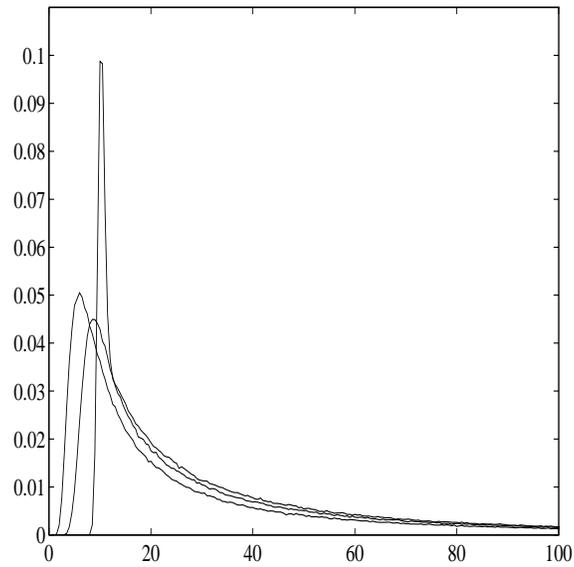}
\caption{\small As in Figure 1, with $\lambda=0.2$, $\mu=0.3$, $c=v=1$,
for $\sigma=1.8, 1, 0.2$ (left to right near the origin).}
}
\end{figure}
%
\begin{figure} 
\centering{
\epsfxsize=7.5cm
\epsfysize=7.5cm
\epsfbox{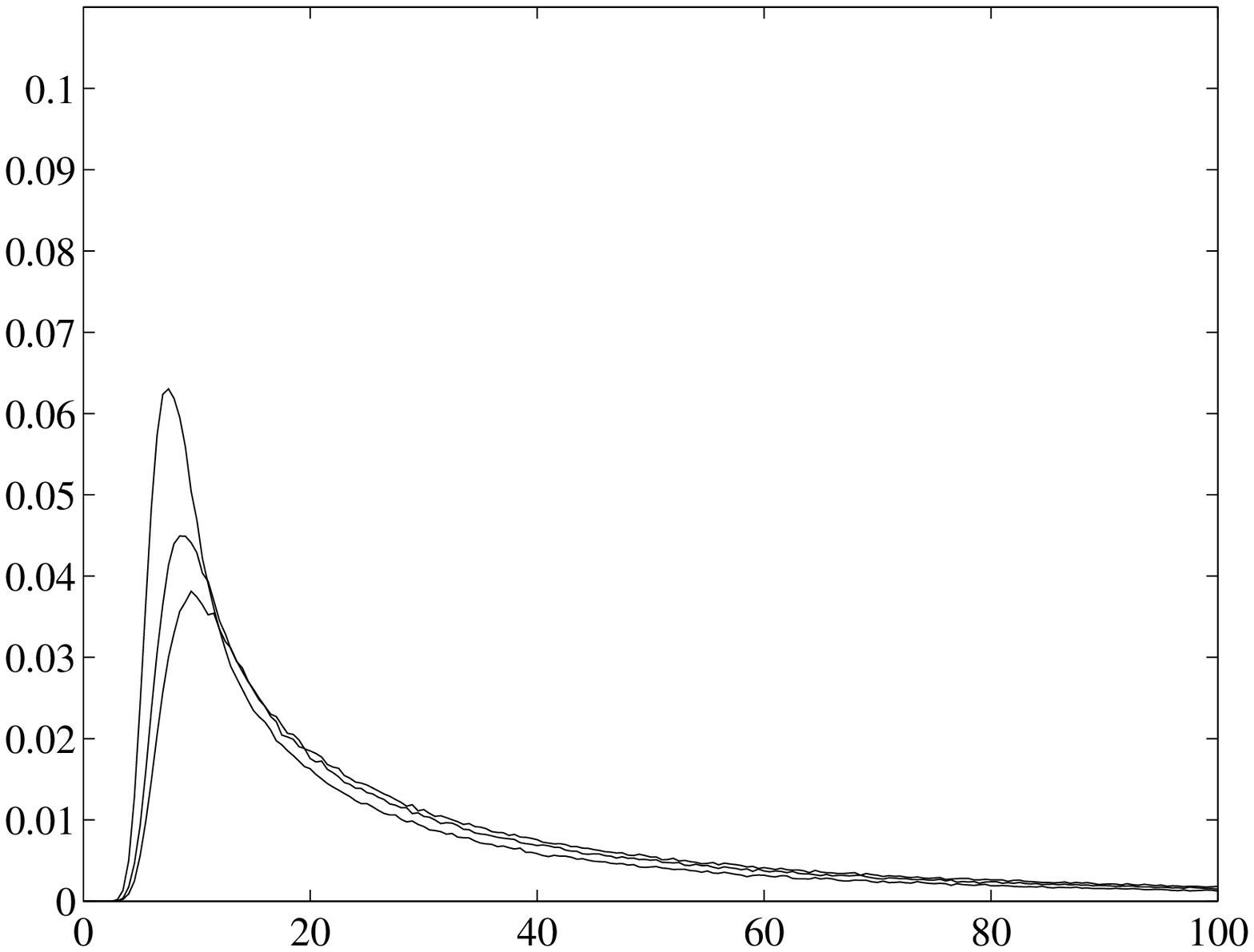}
\caption{\small As in Figure 1, with $\lambda=0.2$, $\mu=0.3$, $\sigma=1$,
for $c=v=0.8, 1, 1.2$ (bottom to top near the origin).}
}
\end{figure}
%
\section{Erlang distributed inversion times}\label{section:5bis}
In this Section we consider the case when the inversion times of the Wiener neuronal 
model~(\ref{equation:1}) have distribution functions of Erlang type with shape 
parameter 2: $F_U(t)=1-\lambda t\,e^{-\lambda t}$ and $F_D(t)=1-\mu t\,e^{-\mu t}$, 
$t\geq 0$. The mean values of the times intervals separating consecutive 
inversions of the drift are now given by $2/\lambda$ and $2/\mu$. The analysis has 
been performed on some cases analogous to those presented in Section~\ref{section:5}, 
with $\mu\geq\lambda$, with threshold $\beta=10$, and with $x_0=0$. 
\par
As for the case of exponential inversion times, the estimated firing densities are 
unimodal (see Figures 5$\div$8). Moreover, if $\mu$ increases the median $q_2$ 
decreases, as well as the quartiles $q_1$ and $q_3$, and the inter-quartile range 
$IQR$ (see Figure 5). A similar behaviour is again observed if $\lambda$ 
decreases (see Figure 6). We emphasize that also in this case the firing density 
becomes less peaked as $\sigma$ increases (see Figure 7). Finally, when $c$ and $v$ 
increase, with $c=v$, then $q_1,q_2,q_3$ decrease (see Figure 8). A similar behaviour 
has also been exhibited by the estimated firing densities in the case of exponentially 
distributed inversion times (compare Figures 5$\div$8 with Figures 1$\div$4). 
\begin{figure} 
\centering{
\epsfxsize=7.5cm
\epsfysize=7.5cm
\epsfbox{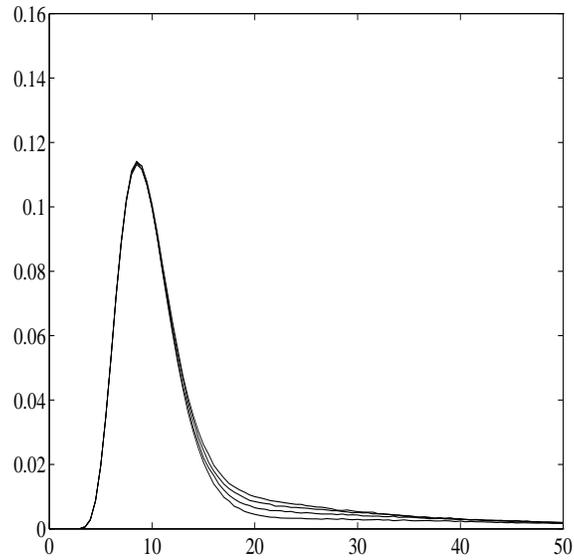}
\caption{\small Estimated firing density of the Wiener model with
Erlang alternating drift for $\sigma=1$, $\lambda=0.1$, $c=v=1$,
for $\mu=0.10, 0.15, 0.20, 0.25$ (bottom to top near abscissa 20).}
}
\end{figure}
%
%
\begin{figure} 
\centering{
\epsfxsize=7.5cm
\epsfysize=7.5cm
\epsfbox{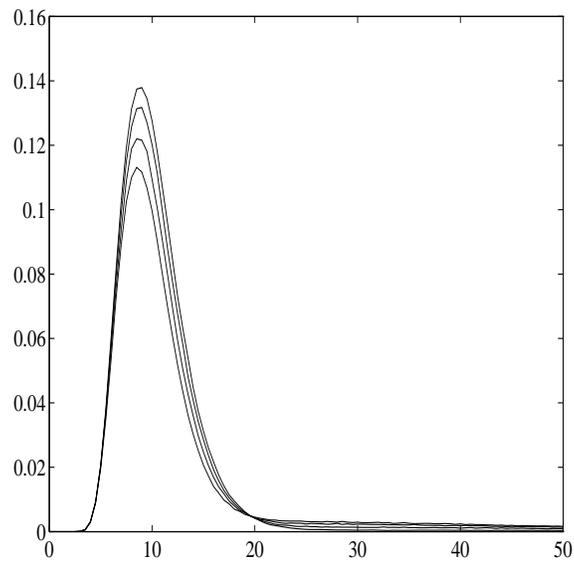}
\caption{\small As in Figure 5, with $\sigma=1$, $\mu=0.1$, $c=v=1$, for $\lambda=0.10, 0.075, 0.05, 0.025$ (bottom to top near the origin).}
}
\end{figure}
%
\begin{figure} 
\centering{
\epsfxsize=7.5cm
\epsfysize=7.5cm
\epsfbox{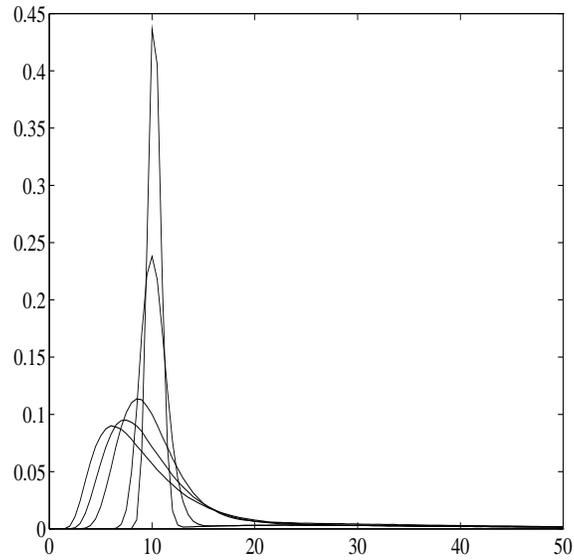}
\caption{\small As in Figure 5, with $\lambda=0.1$, $\mu=0.15$, $c=v=1$,
for $\sigma=1.8,1.4,1.0,0.4,0.2$ (left to right near the origin).}
}
\end{figure}
%
\begin{figure} 
\centering{
\epsfxsize=7.5cm
\epsfysize=7.5cm
\epsfbox{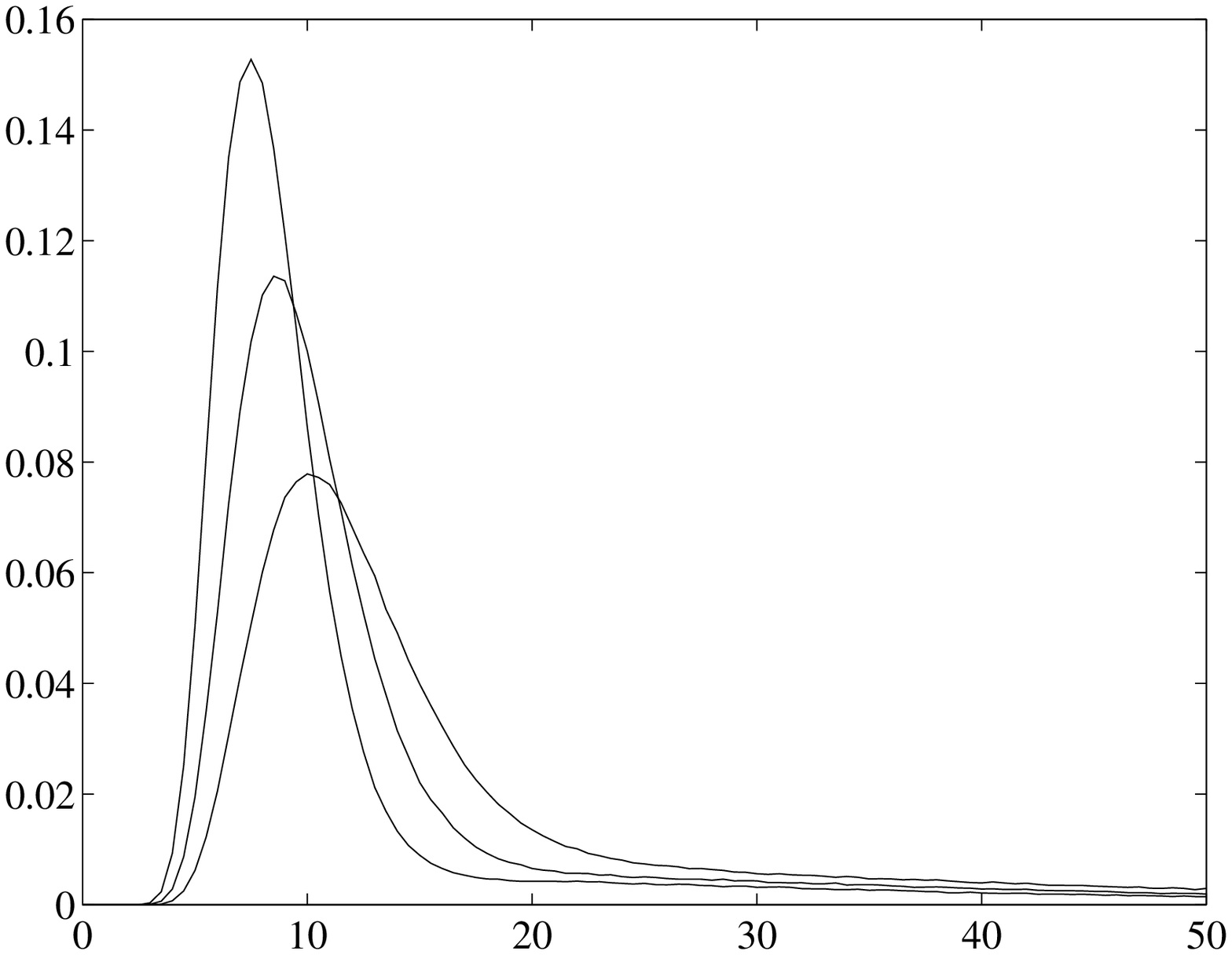}
\caption{\small As in Figure 5, with $\lambda=0.1$, $\mu=0.15$, $\sigma=1$,
for $c=v=0.8, 1, 1.2$ (bottom to top near the origin).}
}
\end{figure}

%
\section{Deterministic inversion times}\label{section:6}
With reference to the model~(\ref{equation:1}) let us assume that $c=v$ and that the 
drift inversion times are deterministic, with $F_U(t)=F_D(t)={\bf 1}_{\{t\geq \tau\}}$. 
This means that the second term at the right-hand-side of~(\ref{equation:1}) 
increases and decreases alternately in a deterministic fashion every $\tau$ 
instants. In other words, 
\begin{equation}
 u(t):=\int_0^t\xi(s)\,{\rm d}s
 \label{equation:4}
\end{equation}
for $t>0$ is a sawtooth function of period $2\tau$. 
\par
As in the previous cases, we have performed $10^6$ simulations to obtain 
estimates of the firing density $g(\beta,t\,|\,x_0)$, with threshold 
$\beta=10$ and initial state $x_0=0$. 
\par
Firing densities now appear to be quite different from the previous cases. Indeed, 
the estimated firing densities are multimodal. In Figure 9 two examples are shown 
where the peaks are located near the maxima of~(\ref{equation:4}), i.e.\ 
near $\tau,3\tau,5\tau,\ldots\;$. 
\begin{figure} 
\centering{
\epsfxsize=7.5cm
\epsfysize=7.5cm
\epsfbox{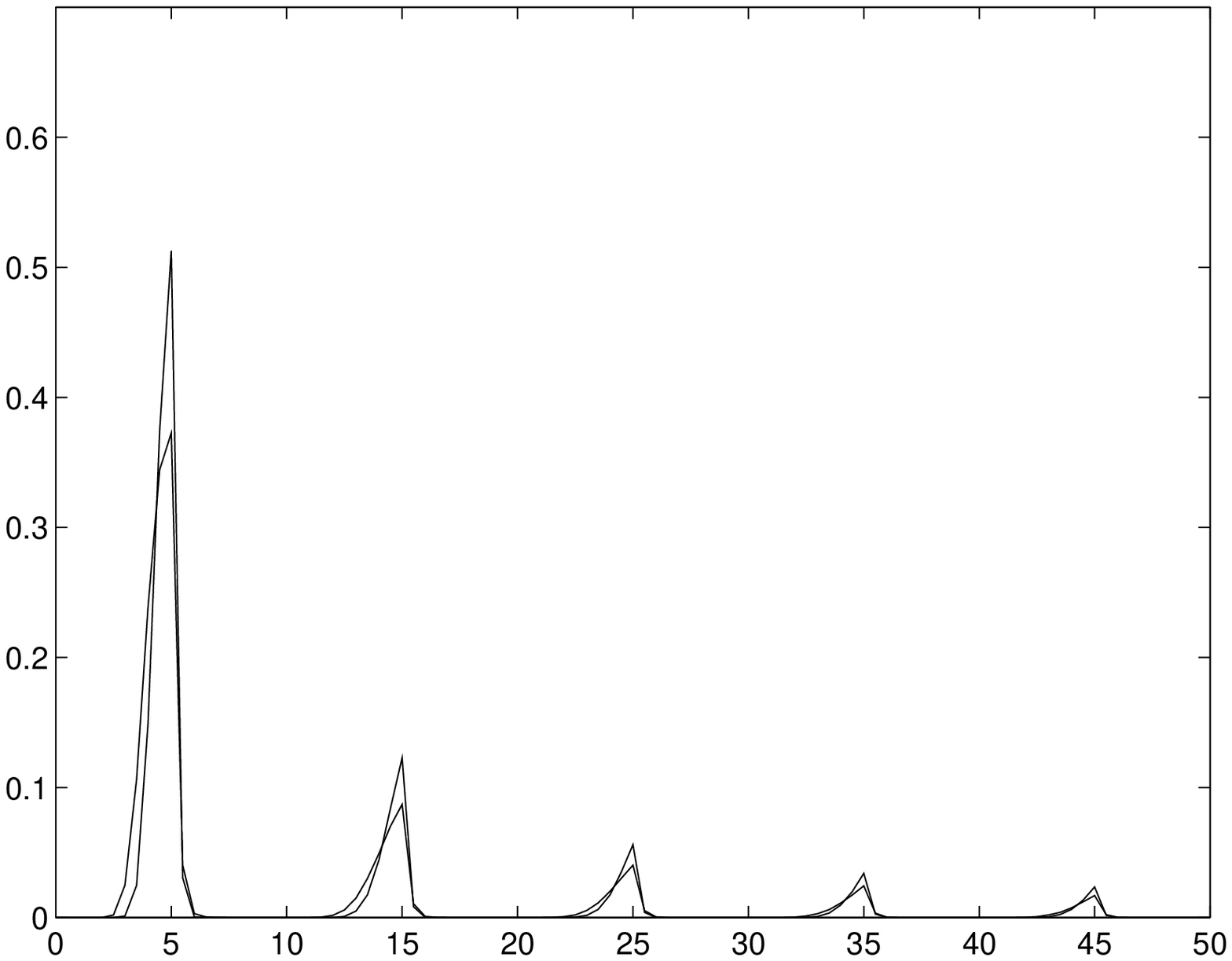}
\caption{\small Estimated firing density of the Wiener model with drift alternating 
at deterministic times for $\tau=5$, $c=v=2$, $\sigma=1$ (lower peaks) and 
$\sigma=0.7$ (higher peaks).}
}
\end{figure}
%
\section{Sinusoidal input}\label{section:7}
In this Section we shall analyse the effect of a smoother input for the
Wiener neuronal model of Section \ref{section:6} by changing function
(\ref{equation:4}) to a smoother one. More specifically, we assume that
the membrane depolarization is now described by the Gauss-Markov diffusion
process $\{\hat X(t);\,t\geq 0\}$ given by
\begin{equation}
 \hat X(t)=x_0+\int_0^t\eta(s)\,{\rm d}s+\sigma\,B(t), \quad t\geq 0,
 \label{equation:3}
\end{equation}
with $x_0\in{\bf R}$, $\sigma>0$, and
$$
 \eta(t)=\alpha\,{\pi\over 2\tau}\,\sin\left({\pi\over \tau}\,t\right),
$$
with $\alpha$ and $\tau$ positive. Note that mean and covariance of
$\hat X(t)$ are given by
\begin{equation}
 E[\hat X(t)]={\alpha\over 2}
 \left[1-\cos\left({\pi\over \tau}\,t\right)\right],
 \quad t\geq 0,
 \label{equation:5}
\end{equation}
and
$$
 {\rm Cov}[\hat X(s),\hat X(t)]= \sigma^2\,\min\{s,t\},
 \quad s,t\geq 0,
$$
respectively. Hence, $\alpha$ is the maximum value attained by the mean of
$\hat X(t)$. To compare the present model with that considered in the previous
section, we choose the involved parameters in such a way that the functions
(\ref{equation:4}) and (\ref{equation:5}) possess equal maxima and minima.
Hence, according to the parameters' values chosen in Figure 9 we set 
$\alpha=c\,\tau$, with $c=v$, $\beta=10$ and $\tau=5$.
\par
To determine the firing density for the model (\ref{equation:3}) we resort 
to a numerical procedure due to Di~Nardo {\em et al.} (2001). 
The results obtained show that the firing densities
are still multimodal (see Figure 10), as for the model described in
Section \ref{section:6}, and that the peaks are located around the maxima
of (\ref{equation:5}). The shapes of the firing densities are similar to those
obtained for the Wiener model with alternating drift, though being now
characterized by less sharp peaks. This is also evident by comparing
Figure 10 with Figure 9. 
\begin{figure} 
\centering{
\epsfxsize=7.5cm
\epsfysize=7.5cm
\epsfbox{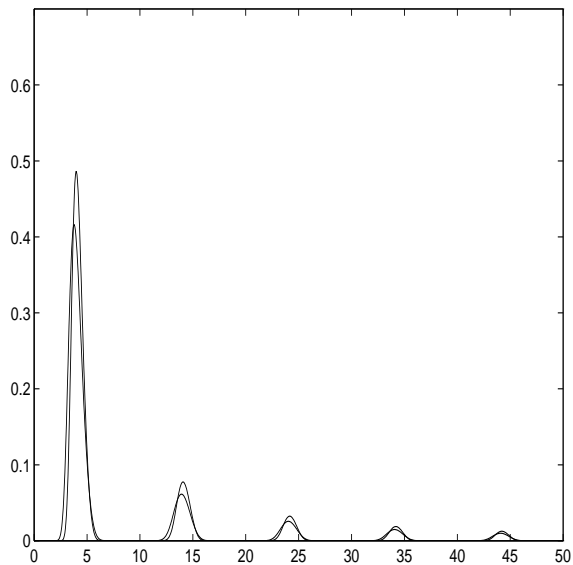}
\caption{\small The firing density of the Wiener model with sinusoidal drift for 
$\tau=5$, $\alpha=10$ $\sigma=1$ (lower peaks) and $\sigma=0.7$ (higher peaks).}
}
\end{figure}
%
\section{Concluding remarks}\label{section:8}
The Wiener neuronal model has been considered by assuming that the effects of 
alternating stimuli are included into the drift. Making use of a numerical 
procedure to simulate sample-paths of the resulting process, we have obtained 
estimates of the firing densities as FPT densities through a constant threshold. 
Three cases have been considered: the time periods separating consecutive
changes of the drift {\em (i)\/} are exponentially distributed, {\em (ii)\/} 
they have an Erlang-type distribution, and {\em (iii)\/} they possess constant 
length. For sensible choices of the involved parameters, the results obtained 
in these three cases are quite different. Indeed, while the presence 
of randomness in the drift of the process produces unimodal firing densities 
in cases {\em (i)\/} and {\em (ii)\/}, the deterministic behaviour of the drift 
yields multimodal densities in case {\em (iii)\/}. The computational results found 
in case {\em (iii)\/} have been finally compared with those holding for the Wiener
neuronal model subject to oscillating input of sinusoidal type. In this case the
firing densities are still multimodal, though less peaked than in case {\em (iii)\/}.
\section*{Acknowledgements}
This work has been performed within a joint cooperation agreement between Japan Science
and Technology Coorporation (JST) and Universit\`a di Napoli Federico II, under partial
support by MIUR (PRIN 2000).

\end{document}